\documentclass[a4paper,12pt]{article}

\newcommand{\sect}[1]{\setcounter{equation}{0}\section{#1}}

\begin{document}
\topmargin 0pt \oddsidemargin 0mm

\renewcommand{\thefootnote}{\fnsymbol{footnote}}
\begin{titlepage}
\begin{flushright}
 hep-th/0403134
\end{flushright}

\vspace{5mm}
\begin{center}
{\Large \bf A Note on Curvature Fluctuation of  Noncommutative
Inflation } \vspace{32mm}

{\large Rong-Gen Cai\footnote{e-mail address: cairg@itp.ac.cn}}

\vspace{10mm} {\em  Institute of Theoretical Physics, Chinese
Academy of Sciences,\\
 P.O. Box 2735, Beijing 100080, China \\
 Department of Physics, Baylor University, Waco, TX76798-7316}

\end{center}

\vspace{5mm} \centerline{{\bf{Abstract}}}
 \vspace{5mm}
An elegant approach, which incorporates the effect of the stringy
spacetime uncertainty relation, to calculate power spectra of
fluctuations during inflation has been suggested by Brandenberger
and Ho. In this approach, one of important features is the
appearance of an upper bound on the comoving momentum $k$, at
which the stringy spacetime uncertainty relation is saturated. As
a result, the time-dependent upper bound  leads us to choose
naturally a set of initial vacua, in which the stringy uncertainty
relation is saturated.  In this note, with that set of vacua we
calculate power spectrum of curvature fluctuation for a power law
inflation, up to the leading order of a parameter describing the
spacetime noncommutativity. It turns out that this choice of
initial vacuum has a significant effect on the power spectrum of
fluctuations.

\end{titlepage}

\newpage
\renewcommand{\thefootnote}{\arabic{footnote}}
\setcounter{footnote}{0} \setcounter{page}{2}

\sect{Introduction}
 String/M theory as a fundamental theory should produce a cosmology model
 describing our universe. However, our understanding so far to the
 string/M theory is quite uncomplete, even we have not yet had a
 successful perturbative string theory in a time-dependent
 background. Although so, it is generically believed that some
 remarkable features, for instance, spacetime uncommutativity,
 of present string/M theory (quantum gravity
theory) will manifest around the string scale in an effective
field theory description.

On the other hand, according to the inflation
model~\cite{Guth,book1,book2}, the large scale structure of the
universe we observed today is formed starting from quantum
fluctuations during inflation; those tiny quantum fluctuations are
stretched to cosmo scales with the expansion of the universe. With
the precisely astronomical observations, now it becomes possible
to observe the effects of spacetime structure around the string
scale in the modern cosmology. Indeed, in recent years some
authors studied the effect of spacetime uncommutativity on
inflation models~\cite{Chu}-\cite{Li}.

In string theory, there is a universal spacetime uncertainty
relation, $\triangle t_p \triangle x_p \ge l_s $, where $t_p$ and
$x_p$ are physical time and distance, and $l_s=M_s$ is the string
length scale. This stringy spacetime uncertainty relation (SSUR)
holds not only for perturbative string theory, but also for
nonperturbative string theory~\cite{YL}. The SSUR implies that at
the string scale spacetime is noncommutative, and  field theory in
such a background must be nonlocal. However, it turns out that it
is not so easy to incorporate the spacetime noncommutativity to
the effective description of field theory in inflation model. In a
recent paper \cite{BH} by Brandenberger and Ho, an elegant
approach is proposed, in which the spacetime noncommutativity is
incorporated through replacing the usual product by star product
of noncommutative field theory in the effective action describing
the fluctuations of inflaton and metric. The effect of spacetime
noncommutativity on the power spectrum of fluctuations with the
Brandenberger-Ho's approach has been investigated in some recent
papers. In particular, papers \cite{HL1,TMB,HL2,HL3} used the
effect to fit the WMAP data~\cite{WMAP}, showing that the
noncommutativity effect could produce a large running of spectrum
index in some inflation models, but it is not enough to generate
the suppression of angle power spectrum at large scales, as the
WMAP data indicate. Myung et al~\cite{Myung1} calculated the power
spectrum of curvature fluctuation up to the second order of slow
roll parameters and noncommutativity parameter.
Following~\cite{BH}, Liu and Li in \cite{Li} discussed the scalar
fluctuations of tachyon inflation in a noncommutative spacetime.

Since the background (the Friedmann-Robertson-Walker spacetime) is
homogeneous and isotropic, in the Brandenberger-Ho's approach, the
spacetime noncommutativity has not any effect on the background
evolution. However, two new features occur in this approach,
compared to inflation model in the commutative (usual) spacetime.
One of them is that the coupling between the fluctuation modes and
the background becomes nonlocal in time. The other is that there
exists a critical time for each mode, at which the SSUR is
saturated, which indicates that the critical time is the one when
the mode is generated, before that the mode does not exist. This
critical time leads to a natural choice of initial vacuum for
given a mode. However, when calculating the power spectrum of
fluctuations, most of existing literature still choose the
adiabatic vacuum as the initial vacuum.  In this note we emphasize
the significance of the natural choice of initial vacuum, and
calculate the power spectrum of curvature fluctuation for a power
law inflation in a noncommutative spacetime. It turns out that
indeed this choice of vacuum has a significant effect on the power
spectrum of fluctuations.

\sect{Curvature perturbation of noncommutative inflation}

The model incorporating the SSUR can be written down as~\cite{BH}
\begin{equation}
\label{2eq1}
 S=V\int_{k<k_0}d\tilde \eta d^3k z^2_k(\tilde \eta)
 (\phi'_{-k} \phi'_k-k^2\phi_{-k}\phi_k),
 \end{equation}
 where $V$ denotes the total spatial volume and $z_k$ is some
 smeared version of $z=a\dot \phi_0/H$ or the scale factor $a$ over a range of time of
 characteristic scale $\triangle \tau=l_s^2 k $ with the overdot being the
 derivative with respect to the cosmic time, $\phi_k$ is the Fourier mode
 with comoving wave number $k$ of scalar perturbation of metric, $\phi_0$ represents
 the background inflaton, and $H$ is the Hubble parameter (for more details, see
 also the appendix in \cite{BH}).  In
 addition,
 \begin{equation}
 \label{2eq2}
 z_k^2= z^2 y_k^2(\tilde \eta), \ \ \ \tilde z_k(\tilde\eta)=a_{\rm
 eff}(\tilde \eta).
 \end{equation}
 Here the modified  conformal time $\tilde \eta$ is defined via
 $d\tilde\eta =\tilde z_k^{-2} d \tau $, the coordinate time $\tau$ is related
 to the cosmic time $t$ via $d\tau = a dt$, and the conformal time $\eta $ is
 defined as usual as $d\eta = a^{-1}dt$. Namely, a flat
 Friedmann-Robertson-Walker metric can be written in the following
 forms in terms of different times
 \begin{eqnarray}
 \label{2eq3}
 ds^2 &=& dt^2 -a^2(t) (dr^2 +r^2 d\Omega^2), \nonumber \\
       &=& a^2(\eta) (d\eta^2 -(dr^2 +r^2 d\Omega^2)), \nonumber \\
       &=& a^{-2}(\tau)d\tau^2 -a^2(\tau) (dr^2 +r^2 d\Omega^2).
       \end{eqnarray}
Finally, $y_k^2$ and $a_{\rm eff}$ in (\ref{2eq2})
 are given by
 \begin{equation}
 y_k=(\beta^-_k \beta^+_k)^{1/4}, \ \ \ a^2_{\rm
 eff}=(\beta^+_k/\beta^-_k)^{1/2},
 \end{equation}
 where
 \begin{equation}
 \label{2eq4}
 \beta^{\pm}(\tau)=\frac{1}{2}(a^{\pm 2}(\tau-l_s^2k) +a^{\pm
 2}(\tau +l_s^2k)).
 \end{equation}
 Note that there exists an upper bound $k_0$ on the wave number in
 the effective action (\ref{2eq1}), which is given by
 \begin{equation}
 \label{2eq6}
 k \le k_0(\tau) =a_{\rm eff}(\tau)/l_s.
 \end{equation}
 This has an important consequence on the perturbation. It implies that given a
 wave number there is a critical time  $\tau_0=\tau_0(k)$.
  At the time when the equation(\ref{2eq6}) holds, the SSUR is
 saturated. This indicates  that before the critical time the mode $k$
 should not exist in order the SSUR to hold.

 The equation of scalar perturbation can be written as
 \begin{equation}
 \label{2eq7}
 \mu _k''+(k^2-\frac{z''_k}{z_k})\mu_k =0,
 \end{equation}
 where a prime denotes the derivative with respect to $\tilde
 \eta$ and $\mu_k=z_k\phi_k$. Defining the slow-roll parameters
 \begin{equation}
 \epsilon =-\frac{\dot H}{H^2}=\frac{1}{2}\left (\frac{\dot \phi_0}{H}\right)^2,
  \ \ \ \delta = \frac{\ddot \phi_0}{H\dot \phi_0},
  \end{equation}
  and introducing a new parameter describing the noncommutativity of
  spacetime~\cite{HL3}
  \begin{equation}
  \mu = \left(\frac{kH}{aM_s^2}\right)^2,
  \end{equation}
  one has
  \begin{eqnarray}
  \frac{z''_k}{z_k} &=& 2(aH)^2\left(1+\frac{5}{2}\epsilon
  -\frac{3}{2}\delta -2 \mu\right), \nonumber \\
    aH &=& -\frac{1}{\tilde \eta}(1+\epsilon +\mu),
\end{eqnarray}
 up to the leading order of $\epsilon$, $\delta$ and $\mu$. The
 equation (\ref{2eq7}) is then changed to
 \begin{equation}
 \label{2eq11}
\mu _k''+(k^2-\frac{\nu^2-1/4}{\tilde\eta^2})\mu_k =0,
\end{equation}
where $\nu =3/2 +2 \epsilon +\delta$. It is interesting to note
that the equation (\ref{2eq11}) has a same form as the one for the
usual case without spacetime noncommutativity and that the only
difference is to replace the conformal time $\eta$ by the modified
one $\tilde\eta$. Therefore once obtaining the solution of the
equation (\ref{2eq11}) and choosing an appropriate initial vacuum
$|0\rangle $, one has the power spectrum of the curvature
fluctuation ${\cal R}=\mu_k/z_k$,
\begin{equation}
\label{2eq12}
 {\cal P}_{\cal R}(k) = \frac{k^3}{2\pi^2}\frac{\langle 0|
 \hat \mu_k^+\hat \mu_k|0\rangle}{z^2_k}.
 \end{equation}
 It is well-known that there does not exist an unique vacuum for quantum fields in
 curved spaces. As a result, the key point is to choose an appropriate
 vacuum, which could describe correctly physics happened in the situation
 under consideration.
 Usually when one deals with quantum fluctuations during inflation,
 the adiabatic vacuum is chosen. However, it is not suitable to
 choose the adiabatic vacuum here since for a given mode $k$, there exists
 a critical time $\tau_0$ given by equation (\ref{2eq6}), before that the
 mode does not exist.

To choose an appropriate initial vacuum for the present situation,
let us recall the approach developed by Danielsson~\cite{Dan} (for
earlier discussion see \cite{Starb}), which is used to discuss the
effect of the transplanckian physics on inflation~\cite{Dan,ACT}.
There is also an upper bound on the wave number, beyond which the
effective field theory description invalidates.  Therefore the
situation for the transplanckian physics is quite similar to the
present case. In order to incorporate the effect of the upper
bound (\ref{2eq6}), we can therefore follow the approach proposed
by Danielsson~\cite{Dan}.

The momentum which is conjugate to $\mu_k$ is
\begin{equation}
\label{2eq13}
 \pi_k= \mu_k' -\frac{z_k'}{z_k}\mu_k.
 \end{equation}
Since the vacuum will be dependent on time, the Heisenberg picture
is the most convenient one to quantize the system, in which
operators evolve with time, but states do not. In terms of time
dependent oscillators, we have
\begin{eqnarray}
\label{2eq14}
 \hat \mu_k(\tilde \eta) &= &\frac{1}{\sqrt{2k}}(\hat a_k(\tilde\eta)
    +\hat a^+_{-k}(\tilde\eta)), \nonumber \\
\hat \pi_k(\tilde\eta) &=& -i\sqrt{\frac{k}{2}}(\hat
a_k(\tilde\eta)
    -\hat a^+_{-k}(\tilde\eta)).
    \end{eqnarray}
The oscillators can be expressed in terms of their values at a
time $\tilde \eta_0=\tilde\eta_0(\tau_0)$ through a Bogoliubov
transformation
\begin{eqnarray}
\label{2eq15}
 \hat a _k(\tilde \eta)&=& u_k(\tilde\eta)\hat a _k(\tilde \eta_0)
            +v_k(\tilde\eta )\hat a ^+_{-k}(\tilde \eta_0),
            \nonumber \\
\hat a ^+_{-k}(\tilde \eta)&=&
  u^*_k(\tilde\eta)\hat a^+ _{-k}(\tilde \eta_0)
            +v^*_k(\tilde\eta )\hat a _{k}(\tilde \eta_0).
            \end{eqnarray}
  Substituting this into (\ref{2eq14}) yields
  \begin{eqnarray}
  \label{2eq17}
  \hat \mu_k(\tilde\eta)&=&f_k(\tilde\eta)\hat a
  _k(\tilde\eta_0)+f^*_k(\tilde\eta)\hat a ^+_{-k}(\tilde\eta_0),
   \nonumber \\
   i\hat\pi_k(\tilde\eta)&=& g_k(\tilde\eta)\hat a_k(\tilde\eta_0)
     -g^*_k(\tilde\eta)\hat a^+_{-k}(\tilde\eta_0),
     \end{eqnarray}
     where
     \begin{equation}
     f_k(\tilde\eta)=\frac{1}{\sqrt{2k}}(u_k(\tilde\eta)+v^*_k(\tilde\eta)),
     \ \ \
     g_k(\tilde\eta)=\sqrt{\frac{k}{2}}(u_k(\tilde\eta)-v^*_k(\tilde\eta)).
     \end{equation}
     Here $f_k(\tilde\eta)$ is a solution of equation
     (\ref{2eq7}), and the normalization condition is
     \begin{equation}
     \label{2eq18}
     |u_k|^2 -|v_k|^2=1.
     \end{equation}
Obviously a reasonable choice for the initial vacuum is
\begin{equation}
\label{2eq19}
 \hat a_k(\tilde\eta_0)|0,\tilde\eta_0\rangle =0,
 \end{equation}
 since the mode $k$ does not exist when $\tilde\eta < \tilde\eta_0
 $ (note that here $\tilde \eta$ is always negative).
 In this case, one can see from the relation (\ref{2eq14}) that the
 field and its conjugate momentum has a simple relation
 \begin{equation}
 \label{2eq20}
 \hat \pi_k(\tilde\eta_0)|0,\tilde\eta_0\rangle= ik \hat
 \mu_k(\tilde \eta_0)|0,\tilde\eta_0\rangle .
 \end{equation}
 This choice of vacuum has a simple physical interpretation~\cite{Dan}. It
 corresponds to a state which minimizes the uncertainty at the
 time $\tilde \eta=\tilde\eta_0$. This is consistent with the
 statement mentioned above that at the time $\tilde\eta_0$ the
 SSUR is saturated. Also it in turn justifies the choice (\ref{2eq19}) of
 initial vacuum.

For the equation (\ref{2eq14}), one has the following solution
\begin{equation}
\label{2eq21}
 f_k=A_k\sqrt{-\tilde
 \eta}J_\nu(-k\tilde\eta)+B_k\sqrt{-\tilde\eta}Y_{\nu}(-k\tilde\eta),
 \end{equation}
 where $J_{\nu}$ and $Y_{\nu}$ are Bessel functions of the first
 and second kind, respectively, and $A_k$ and $B_k$ are two complex constants.

 Now we consider a power-law inflation with a scale factor
 $a=a_0t^p $, which could be produced by a scalar field with an
 exponential potential $ V=V_0 \exp(-\sqrt{2/p}\phi_0/M_p)$. In
 this case, $\dot \phi_0=\sqrt{2p}M_p/t $, $H=p/t  $, and
 $z=a\sqrt{2/p}M_p$. Here $M_p$ denotes the induced Planck mass.
 In the slow roll approximation, for the power law inflation, one
 has $\epsilon =-\delta =1/p$, so $\nu =3/2 +1/p$. In fact, for
 the power law inflation, the slow roll approximation is not
 necessary, the exact value of $\nu$ is $3/2 +1/(p-1)$. It is seen
 that the slow roll approximation is quite well if $p\gg 1$.
 Therefore, in what follows we will take $\nu =3/2 +1/(p-1)$.
 For the solution (\ref{2eq21}) we have
 \begin{equation}
 \label{2eq22}
 g_k= -i k \sqrt{-\tilde\eta}(A_kJ_{\nu-1}(- k\tilde\eta)
 +B_kY_{\nu-1}(-k\tilde\eta)).
 \end{equation}
Using (\ref{2eq17}) we obtain
\begin{eqnarray}
\label{2eq23}
 u_k &=& \sqrt{-k\tilde\eta/2}
 [A_kJ_{\nu}(-k\tilde\eta)+B_kY_{\nu}(-k\tilde\eta)
  -i( A_k J_{\nu-1}(-k\tilde\eta)+B_kY_{\nu-1}(-k\tilde\eta))],
    \nonumber \\
  v_k^*&=& \sqrt{-k\tilde\eta/2}[A_kJ_{\nu}(-k\tilde\eta)+B_kY_{\nu}(-k\tilde\eta)
  +i( A_k J_{\nu-1}(-k\tilde\eta)+B_kY_{\nu-1}(-k\tilde\eta))].
  \end{eqnarray}
The normalization condition (\ref{2eq18}) yields
\begin{equation}
\label{2eq24}
  A_kB_k^* -A^*_kB_k=-i\frac{\pi}{2}.
  \end{equation}
The vacuum choice (\ref{2eq19}) together with the Bogoliubov
transformation (\ref{2eq15}) implies that $v^*_k(\tilde\eta_0)=0$.
 From (\ref{2eq23}) we then have
\begin{equation}
\label{2eq25}
 A_k= -\frac{Y_{\nu}(-k\tilde\eta_0) +iY_{\nu-1}(-k\tilde\eta_0)}
     {J_{\nu}(-k\tilde\eta_0) +i J_{\nu-1}(-k\tilde\eta_0)}B_k.
\end{equation}
Equations (\ref{2eq24}) and (\ref{2eq25}) determine those two
constants
\begin{eqnarray}
|A_k|^2 &=& -\frac{\pi^2}{8}k\tilde\eta_0
(Y_{\nu}^2(-k\tilde\eta_0) +Y_{\nu-1}^2(-k\tilde\eta_0)),
\nonumber \\
|B_k|^2 &=& -\frac{\pi^2}{8}k\tilde\eta_0
(J^2_{\nu}(-k\tilde\eta_0)+J^2_{\nu-1}(-k\tilde\eta_0)), \nonumber \\
A_kB_k^* &=&-i\frac{\pi}{4}+
\frac{\pi^2}{8}k\tilde\eta_0(J_{\nu}(-k\tilde\eta_0)Y_{\nu}(-k\tilde
  \eta_0)+J_{\nu-1}(-k\tilde\eta_0)Y_{\nu-1}(-k\tilde\eta_0)),
  \nonumber \\
A_k^*B_k &=&i\frac{\pi}{4}+
\frac{\pi^2}{8}k\tilde\eta_0(J_{\nu}(-k\tilde\eta_0)Y_{\nu}(-k\tilde
  \eta_0)+J_{\nu-1}(-k\tilde\eta_0)Y_{\nu-1}(-k\tilde\eta_0)).
  \end{eqnarray}
  Substituting these into (\ref{2eq12}), we obtain the exact power
  spectrum of the curvature fluctuation
  \begin{eqnarray}
  \label{2eq27}
  {\cal P}_{\cal R}(k)&=&\frac{k^4\tilde\eta_0\tilde\eta }{16z^2
  y_k^2(\tilde\eta)}[(Y^2_{\nu}(-k\tilde\eta_0)+Y^2_{\nu-1}(-k\tilde\eta_0))J^2_{\nu}(-k\tilde\eta)
    \nonumber \\
         && +(J^2_{\nu}(-k\tilde\eta_0)+J^2_{\nu-1}(-k\tilde\eta_0))Y^2_{\nu}(-k\tilde\eta)
          \nonumber \\
          &&-
          2 (J_{\nu}(-k\tilde\eta_0)Y_{\nu}(-k \tilde\eta_0)
          +J_{\nu-1}(-k\tilde\eta_0)Y_{\nu-1}(-k\tilde\eta_0))J_{\nu}(-k\tilde\eta)Y_{\nu}
          (-k\tilde\eta)].
          \end{eqnarray}
If one does not consider the effect of spacetime noncommutativity,
replacing $\tilde \eta \to \eta$ and setting $y_k=1$, one can then
obtain from (\ref{2eq27}) the spectrum of curvature perturbation
for a power-law inflation taking into account the transplanckian
physics~\cite{ACT}. Note that the authors of \cite{ACT} only give
the spectrum for a massless scalar field for a power law
inflation. Therefore compared to the case of commutative
spacetime, due to the spacetime noncommutativity, two new features
appear in (\ref{2eq27}): (1) the modified conformal time
$\tilde\eta$ replaces the usual conformal time $\eta$; (2) a new
factor  $y_k^2$ occurs in (\ref{2eq27}). if considering
$\tilde\eta_0 \to -\infty$ and $\tilde\eta \to 0^-$, one is then
led to the spectrum obtained in the adiabatic vacuum.

 Strictly specking, the power spectrum should be calculated
when the mode $k$ crosses the Hubble horizon. That is,
\begin{equation}
\label{2eq28}
 k^2=z''_k/z_k. \end{equation}
 From (\ref{2eq11}), we can see that
it corresponds to $\tilde\eta=-\sqrt{\nu^2-1/4}/k$. The initial
time $\tilde\eta_0$ is related to the coordinate time $\tau_0$ via
$\tilde\eta_0=\tilde\eta_0(\tau_0)$, while the latter is given via
the equation (\ref{2eq6}). For a power law inflation, it is
\cite{BH}
\begin{equation}
\label{2eq29}
 \tau_0=
\left(\left(\frac{kl_s}{\alpha_0}\right)^{2(p+1)/p}+k^2l_s^4\right)^{1/2},
\end{equation}
where $\alpha_0=((p+1)^pa_0)^{1/(p+1)}$. The scale factor has the
form $a=\alpha_0 \tau^{p/(p+1)}$ in terms of the time $\tau$.
Further, the modified conformal time $\tilde \eta $ has the
relation to the coordinate time $\tau$,
\begin{equation}
\label{2eq30}
 \tilde\eta=
-\left(1+\frac{p+1}{p}\mu\right)\frac{p+1}{p-1}\frac{1}{\alpha^2_0\tau^{(p-1)/(p+1)}},
\end{equation}
where
\begin{equation}
\label{2eq31}
 \mu = \frac{p^2}{(p+1)^2}
\left(\frac{kl_s^2}{\tau}\right)^2. \end{equation}
 Therefore the
power spectrum of curvature fluctuation is
 \begin{eqnarray}
 \label{2eq32}
 {\cal P}_{\cal R}(k)&=&\frac{k^{3-2p/(p-1)}p}{32
 M_p^2\alpha_0^6
 \tau_0^{(p-1)/(p+1)}}
  \left (\alpha_0^2\sqrt{\nu^2-1/4}\right )^{(3p-1)/(p-1)}
  \nonumber \\
  && \times
    \left (\frac{p-1}{p+1}\right)^{(p+1)/(p-1)}
    \left(1-\frac{4p}{p-1}\mu_k \right )
    \left (1+\frac{p+1}{p}\mu_0\right)
    \nonumber \\
         && \times [(Y^2_{\nu}(-k\tilde\eta_0)+Y^2_{\nu-1}(-k\tilde\eta_0))J^2_{\nu}(\sqrt{\nu^2-1/4})
         \nonumber \\
         && +(J^2_{\nu}(-k\tilde\eta_0)+J^2_{\nu-1}(-k\tilde\eta_0))Y^2_{\nu}(\sqrt{\nu^2-1/4})
          \nonumber \\
          && - 2 (J_{\nu}(-k\tilde\eta_0)Y_{\nu}(-k \tilde\eta_0)
          +J_{\nu-1}(-k\tilde\eta_0)Y_{\nu-1}(-k\tilde\eta_0))
          \nonumber \\
          && \times
          J_{\nu}(\sqrt{\nu^2-1/4})Y_{\nu}
          (\sqrt{\nu^2-1/4})].
          \end{eqnarray}
 Here
 \begin{equation}
 \mu_0 = \frac{p^2}{(p+1)^2} \left(\frac{kl_s^2}{\tau_0}\right)^2, \ \
  \mu_k=\frac{p^2}{(p+1)^2} \left(\frac{kl_s^2}{\tau_k}\right)^2,
  \end{equation}
  and
  \begin{equation}
  \tau_k=
  \left(\frac{p+1}{p-1}\right)^{(p+1)/(p-1)}\left(\frac{k}{\alpha_0^2
  \sqrt{\nu^2-1/4}}\right)^{(p+1)/(p-1)}.
  \end{equation}
  Now we can discuss two limits, namely the UV region and IR
  region. In the UV region, the first term in (\ref{2eq29}) is
  dominant. In this case, the power spectrum can be approximately
  expressed as
  \begin{eqnarray}
  \label{2eq35}
{\cal P}_{\cal R}(k)&=& \frac{k^{-\frac{2}{p-1}+\frac{1}{p}} p
\alpha_0^{\frac{4}{p-1}}}{32M_p^2}
   \left(\frac{\alpha_0}{l_s}\right)^{\frac{p-1}{p}}
   \left(\nu^2-\frac{1}{4}\right)^{\frac{3p-1}{2(p+1)}}
   \left(\frac{p-1}{p+1}\right)^{(p+1)/(p-1)}
   \nonumber \\
   && \times \left(1+\frac{\alpha_0^2
   l_s^2}{2}\left(\frac{\alpha_0}{kl_s}\right)^{2/p}-\frac{4p}{p-1}\mu_k\right)
   \nonumber \\
   && \times
   [(Y^2_{\nu}(-k\tilde\eta_0)+Y^2_{\nu-1}(-k\tilde\eta_0))J^2_{\nu}(\sqrt{\nu^2-1/4})
         \nonumber \\
         && +(J^2_{\nu}(-k\tilde\eta_0)+J^2_{\nu-1}(-k\tilde\eta_0))Y^2_{\nu}(\sqrt{\nu^2-1/4})
          \nonumber \\
          && - 2 (J_{\nu}(-k\tilde\eta_0)Y_{\nu}(-k \tilde\eta_0)
          +J_{\nu-1}(-k\tilde\eta_0)Y_{\nu-1}(-k\tilde\eta_0))
          \nonumber \\
          && \times
          J_{\nu}(\sqrt{\nu^2-1/4})Y_{\nu}
          (\sqrt{\nu^2-1/4})],
          \end{eqnarray}
   where
   \begin{equation}
   -k\tilde\eta_0=\frac{p+1}{p-1}\frac{1}{\alpha_0
   l_s}\left(\frac{kl_s}{\alpha_0}\right)^{1/p}.
   \end{equation}
In the adiabatic vacuum, which corresponds to the case
$\tilde\eta_0 \to -\infty$,  the spectrum has a form $ {\cal
P}_{\cal R}(k)\sim k^{-2/(p-1)}$ with spectrum index $n_s
=1-2/(p-1)$, up to the leading order.  Now for the Danielsson
vacuum (\ref{2eq19}), it becomes $ {\cal P}_{\cal R}(k) \sim
k^{-2/(p-1) +1/p}[(J_{\nu}(y)Y_{\mu}(x) -Y_{\nu}(y)J_{\nu}(x))^2
+(J_{\nu-1}(y)Y_{\nu}(x) -Y_{\nu-1}(y)J_{\nu}(x))^2]$, where
$y=-k\tilde\eta_0$ and $x=\sqrt{\nu^2-1/4}$. In this case, the
spectrum index will be larger than the one for the adiabatic
vacuum.

On the other hand, in the IR region, the second term in
(\ref{2eq29}) is dominant. In that case, the power spectrum is
approximated to
\begin{eqnarray}
\label{2eq37}
{\cal P}_{\cal R}(k)&=&
   \frac{k^{\frac{2}{p+1}-\frac{2}{p-1}}p
   \alpha_0^{\frac{4}{p-1}}}{32M_p^2l_s^{2(p-1)/(p+1)}}
   \left(\nu^2-\frac{1}{4}\right)^{\frac{3p-1}{2(p-1)}}
   \left(\frac{p-1}{p+1}\right)^{(p+1)/(p-1)}
   \nonumber \\
   && \times \left(1+\frac{p}{p+1}-\frac{4p}{p-1}\mu_k\right)
   \nonumber \\
  && \times
   [(Y^2_{\nu}(-k\tilde\eta_0)+Y^2_{\nu-1}(-k\tilde\eta_0))J^2_{\nu}(\sqrt{\nu^2-1/4})
         \nonumber \\
         && +(J^2_{\nu}(-k\tilde\eta_0)+J^2_{\nu-1}(-k\tilde\eta_0))Y^2_{\nu}(\sqrt{\nu^2-1/4})
          \nonumber \\
          && - 2 (J_{\nu}(-k\tilde\eta_0)Y_{\nu}(-k \tilde\eta_0)
          +J_{\nu-1}(-k\tilde\eta_0)Y_{\nu-1}(-k\tilde\eta_0))
          \nonumber \\
          && \times
          J_{\nu}(\sqrt{\nu^2-1/4})Y_{\nu}
          (\sqrt{\nu^2-1/4})],
\end{eqnarray}
where
\begin{equation}
-k\tilde\eta_0=
\frac{p+1}{p-1}\frac{(kl_s)^{2/(p+1)}}{\alpha_0^2l_s^2}.
\end{equation}
Here one can see that the spectrum index increases, compared to
the case of the UV region. The Danielsson vacuum enhances the blue
tilt.

 An extremal IR limit is  that
$\tilde\eta \to \tilde\eta_0$. This implies that the mode is
generated outside the Hubble horizon; once generated, it becomes
classical fluctuation. In this case, one has from (\ref{2eq27}),
\begin{equation}
\label{2eq39}
  {\cal P}_{\cal R}(k)=\frac{k^{4/(p+1)}p}{2^{(4p+2)/(p+1)}\pi^2
  M_p^2\alpha_0^4
  l_s^{(6p-2)/(p+1)}}\left( 1-\frac{p}{p+1}
  \frac{5}{4\alpha_0^2l_s^2}
  \left(\frac{kl_s}{\alpha_0}\right)^{2/p}\right),
  \end{equation}
which is same as the one obtained in \cite{HL2}. From
(\ref{2eq39}), one can see that the spectrum index $n_s=1
+4/(p+1)$ and the spectrum becomes blue tilt. Note that for modes
generated inside the horizon, the spectrum is  always red tilt
with $n_s=1-2/(p-1)$ if without the spacetime noncommutativity.
The noncommutativity leads a blue tilt spectrum. The Danielsson
vacuum enhances the trend from the red tilt to blue tilt, which
can be seen from (\ref{2eq35}) and (\ref{2eq37}). The comoving
wave number when the spectrum is from red tilt to blue tilt can be
determined by studying (\ref{2eq32}). This needs further analysis
and numerical calculation, which  are currently under
investigation.

In addition, sometimes people calculate the spectrum at the end of
inflation (superhorizon scale), namely, $\tilde \eta \to 0^-$. In
this case, $Y_{\nu}$ dominates over $J_{\nu}$. Up to the leading
order, the spectrum can be simplified to
\begin{eqnarray}
 {\cal P}_{\cal R}(k) &=&-
 \frac{k^{1-2/(p-1)}p \Gamma^2(\nu) \alpha_0^{2(p+1)/(p-1)}\tilde\eta_0}
 {2^{2-2/(p-1)}\pi^2 M_p^2 }
 \left(\frac{p-1}{p+1}\right)^{2p/(p-1)}
 \nonumber \\
 && \times
 [J^2_{\nu}(-k\tilde\eta_0)+J^2_{\nu-1}(-k\tilde\eta_0)],
 \end{eqnarray}
 where $\tau_0$ is given by (\ref{2eq29}) and $-k \tilde\eta_0$ is given by
 (\ref{2eq30}). If one further takes  limit $ k\tilde\eta_0 \to -\infty $ and
 expands to leading order for small $k$, one has
 \begin{eqnarray}
 \label{2eq41}
{\cal P}_{\cal R}(k) &=&
 \frac{k^{-2/(p-1)}p\Gamma^2[3/2
+1/(p-1)]\alpha_0^{2(p+1)/(p-1)}}{2^{(p-3)/(p-1)}\pi^3 M_p^2}
\left(\frac{p-1}{p+1}\right)^{2p/{(p-1)}}
  \nonumber \\
  && \times
 \left(1 +\frac{P_0}{H_0}\sin \left(\frac{2p}{p-1}\frac{P_0}{H_0}-
  \frac{\pi p}{p-1}\right)\right),
  \end{eqnarray}
  where
  \begin{equation}
  P_0= \frac{k}{\alpha_0 (kl_s^2)^{p/(p+1)}}, \ \ \  H_0
  =\frac{\alpha_0p}{p+1}\left(kl_s^2\right)^{-1/(p+1)},
  \end{equation}
  they are physical momentum of mode $k$ and the Hubble parameter
  at the time $\tau_0= kl_s^2$, respectively. Clearly the factor in the second line
  of (\ref{2eq41}) is the effect of the ``transplanckian
  physics''~\cite{Dan,ACT}.

\sect{Conclusions}

Studying spacetime structure around a fundamental scale (planck
scale or string scale) is an interesting issue in its own right.
The present form of string/M theory tells us that spacetime is
noncommutative around the string scale. Naturally it is
interesting to ask whether there are observable effects of the
spacetime noncommutativity in experiments. The precision
observations to the microwave background radiation of the universe
make possible to see the effect. On the other hand, observation
data can in turn also give a constraint on the string scale. In
the approach suggested by Brandenberger and Ho~\cite{BH}, the
spacetime noncommutativity causes two new and interesting features
in calculating the power spectrum of fluctuations during
inflation. In this note we stressed the consequence of existence
of a critical time for a given mode. The critical time leads to a
natural choice of initial vacuum. Following the approach developed
by Danielsson~\cite{Dan,ACT}, we calculated the power spectrum of
curvature fluctuation for a power law inflation. It turns out that
the choice of initial vacuum has a significant effect on the power
spectrum. Compared to the results given in \cite{HL1,TMB,HL2},
which are obtained in the case of the adiabatic vacuum, from our
results (\ref{2eq35}) and (\ref{2eq37}), it can be derived that
the spectrum index increases and the running of spectrum index
becomes larger, which is welcome by the WMAP data.  For modes
generated inside the Hubble horizon, the spectrum is red tilt,
$n_s=1-2/(p-1)$ in the adiabatic vacuum (The spectrum is always
red tilt if without the spacetime noncommutativity). In the
extremal IR limit, where the mode is generated outside the Hubble
horizon, the spectrum is blue tilt, $n_s=1+4/(p+1)$ [see
(\ref{2eq39})]. In the Danielsson vacuum, the spectrum index
increases, compared to the adiabatic vacuum, it is favor to the
flow of spectrum from red tilt to blue tilt. The differences of
spectrum index and its running between the Danielsson vacuum and
adiabatic vacuum have to be numerically determined by
(\ref{2eq32}).

\section*{Acknowledgments}
The author would like to thank Qing-Guo Huang, Miao Li and Bin
Wang for useful discussions, and to express his gratitude to the
Physics Department, Baylor University for its hospitality. This
work was supported in part by a grant from Chinese Academy of
Sciences, a grant No. 10325525  from NSFC, and by the Ministry of
Science and Technology of China under grant No. TG1999075401.


\end{document}